\documentclass[11pt]{article}
\usepackage{cite}
\usepackage{amsmath,amsfonts,amssymb}
\usepackage[small,bf,hang]{caption}

\def\hybrid{
        \topmargin -20pt
        \oddsidemargin 0pt
        \headheight 0pt \headsep 0pt
        \textwidth 6.25in 
        \textheight 9.5in 
        \marginparwidth .875in
        \parskip 5pt plus 1pt \jot = 1.5ex}

\hybrid

 \csname
@addtoreset\endcsname{equation}{section}

\def\moth{\mathsurround=0pt}
\newdimen\zo \zo=0pt

\def\tick{\leaders\hrule height 0.5ex depth 0pt \hskip 0.5pt}
\def\upboxfill{$\moth \setbox\zo\hbox{\tick}%
  \hskip 3pt\hbox to 0pt{$\tick$\hss}\hrulefill \hbox to 7.5pt{$\tick$\hss}$}

\def\dtick{\leaders\hrule height .34pt depth 0.5ex \hskip 0.5pt}
\def\downboxfill{$\moth \setbox\zo\hbox{\dtick}%
  \hskip 2pt\hbox to 0pt{$\dtick$\hss}\hrulefill \hbox to 2pt{$\dtick$\hss}$}

\def\bec{\begin{center}}
\def\ec{\end{center}}

\def\be{\begin{equation}}
\def\ee{\end{equation}}
\def\bea{\begin{eqnarray}}
\def\eea{\end{eqnarray}}
\def\ba{\begin{array}}
\def\ea{\end{array}}

\begin{document}

\begin{titlepage}
\begin{center}
\vskip 2.5cm
{\Large \bf {
Matrix model correlators from non-Abelian T-dual of $AdS_5 \times S^5 $}}\\
\vskip 1.0cm
{\large {Dibakar Roychowdhury}}
\vskip 1cm
{\it {Department of Physics}}\\
{\it {Indian Institute of Technology Roorkee}}\\
{\it {Roorkee 247667, Uttarakhand, India}}\\

\vskip 2.5cm
{\bf Abstract}
\end{center}

\vskip 0.1cm

\noindent
\begin{narrower}
We study various perturbations and their holographic interpretation for non-Abelian T-dual of $ AdS_5 \times S^5 $ where the T-duality is applied along the $ SU(2) $ of $ AdS_5 $. This paper focuses on two types of perturbations, namely the scalar and the vector fields on NATD of $ AdS_5 \times S^5 $.  For scalar perturbations, the corresponding solutions could be categorised into two classes. For one of these classes of solutions, we build up the associated holographic dictionary where the asymptotic radial mode sources scalar operators for the $ (0+1) $d matrix model. These scalar operators correspond to either a marginal or an irrelevant deformation of the dual matrix model at strong coupling. We calculate the two point correlation between these scalar operators and explore their high as well as low frequency behaviour. We also discuss the completion of these geometries by setting an upper cut-off along the holographic axis and discuss the corresponding corrections to the scalar correlators in the dual matrix model. Finally, we extend our results for vector perturbations where we obtain asymptotic solutions for a particular class of modes. These are further used to calculate the boundary charge density at finite chemical potential.  
\end{narrower}

\end{titlepage}

\newpage

\tableofcontents
\baselineskip=16pt
\section{Introduction and Overview}
The $AdS_5/CFT_4$ correspondence \cite{Maldacena:1997re}, that was originally proposed due to Maldacena, stands as one of the corner stones of the modern theoretical physics with widest possible applications including almost all areas of physics\footnote{See, \cite{Aharony:1999ti}-\cite{Natsuume:2014sfa} for a nice set of reviews on the subject. }. The correspondence \cite{Maldacena:1997re} offers a natural framework that allows us to calculate $ n $-point correlations in strongly coupled gauge theories in the Large N limit. These calculations were first performed in the Euclidean signature \cite{Gubser:1998bc}-\cite{Witten:1998qj} which were then subsequently extended to Minkowski signature in \cite{Son:2002sd}. The general idea behind these calculations stems from the fact that the scalar \cite{Skenderis:2002wp}-\cite{deHaro:2000vlm} or vector perturbations \cite{Kovtun:2008kx} in the bulk $AdS_5$ source operators in the dual $CFT_4$ ($\mathcal{N}=4$ SYM) at strong coupling. The purpose of the present paper is to reinvestigate the above ideas in the context of recently proposed type IIA background \cite{Lozano:2017ole} that is obtained following a non-Abelian T-duality (NATD) \cite{delaOssa:1992vci} inside $AdS_5$.

Global aspects of NATD, as it stands right from the beginning \cite{Alvarez:1993qi}-\cite{Alvarez:1994np}, was unclear from the perspective of the string sigma model. These studies were confined to the NS sector for a long time, until the RR sector was introduced by authors in \cite{Sfetsos:2010uq}. This opens up a new realm for the AdS/CFT correspondence and results in a series of papers \cite{Lozano:2011kb}-\cite{Roychowdhury:2023lxk}. When applied inside $ S^5 $, NATD results in a novel class of type IIA backgrounds \cite{Gaiotto:2009gz} that are dual to $ \mathcal{N}=2 $ super-conformal quivers \cite{Gaiotto:2009we} in four dimensions. However, when NATD is performed using $ SU(2) $ inside $ AdS_5 $, one generates a new class of type IIA background \cite{Lozano:2017ole} that are conjectured to be dual to irrelevant deformations of ($0+1$) dimensional matrix models preserving $\mathcal{N}=2$ SUSY. These are close cousins of BMN Plane Wave Matrix Model (PWMM) \cite{Berenstein:2002jq}-\cite{Maldacena:2002rb}
 which corresponds to massive deformations of BFSS matrix model \cite{Banks:1996vh}.

Following the standard AdS/CFT folklore \cite{Ramallo:2013bua}-\cite{Natsuume:2014sfa}, we claim that scalar field ($\phi$) propagating over NATD $ AdS_5 $ sources some dual operator ($\mathcal{O}$) in the PWMM. The purpose of the present paper is to learn about these dual operators and compute the associated two point correlations by studying the dynamics of scalar field in the dual NATD $ AdS_5 $ background \cite{Lozano:2017ole}.

We look into this ($ 0+1 $)d matrix model as a lower dimensional QFT derived from some higher dimensional parent QFT. The role of NATD is to impose a ``defect'' into this lower dimensional theory which appears by virtue of the irrelevant operators of the theory. As a result, the irrelevant deformation triggers a flow that takes the theory from some lower dimensional fixed point to its higher dimensional parent theory. In the bulk, the irrelevant deformation (or the defect) appears through the presence of the holographic ($ \rho $ or $ \eta $) direction which breaks the original isometries of the $ AdS_5 $. As we show, there are two classes of scalar modes - one of which we classify as ``zero'' modes that are naturally decoupled from the holographic direction and could be given a holographic interpretation in terms of excitations of a ($0+1$) dimensional QFT.
However, these zero modes trigger a class of ``irrelevant'' operators in the dual matrix model which indicates that the system is driven from a lower dimensional QFT to its higher dimensional parent theory. The other class of (non-zero) modes appear explicitly through the holographic direction and therefore carry the ``memory'' of the defect in the lower dimensional QFT or the matrix model. The holographic interpretation for these non-zero modes appears to be rather obscure as the boundary action does not seem to be renormalised. 

We further bound the holographic axis by imposing an upper cut-off ($ \rho_0 \sim L $) with an aim that we only focus upon the lower dimensional theory without concerning about the higher dimensional parent QFT. The resulting procedure yields a $ 1/L^2 $ correction to the two point correlation calculated for the NATD background. We call these corrections (at leading order) as ``improvement'' terms that arise as an artefact of the finiteness of the holographic direction.

The analysis for vector modes follows in an identical fashion where we classify the zero and the non-zero modes like in the case of scalars. The zero modes are as usual decoupled from the holographic $\eta$- direction and thereby represent excitations of a defect QFT. Non-zero modes, on the other hand, are coupled to the holographic $\eta$- axis, which thereby carry the information about the defect in the ($0+1$)d field theory. We build up the holographic dictionary corresponding to these zero modes where we construct the renormalised action for the boundary theory and comment on the corresponding charge-charge correlation at finite density.

The organisation for the rest of the paper is as follows. In Section 2, we discuss dynamics of scalar perturbations over NATD $AdS_5$ where we build up the associated holographic dictionary for a particular class of modes that source operators in the dual QFT$_1$ at strong coupling. We obtain the renormalised boundary action and compute the corresponding two point correlation between the scalar operators in the dual QFT$_1$ and explore their high and low frequency behaviour. In the next Section 3, we estimate $1/L^2$ corrections to these scalar correlators by considering coarse-grained approximations \cite{Lozano:2017ole} of NATD background. In Section 4, we generalise our results for vector perturbations. We classify vector perturbations and build up the associated holographic dictionary for a particular class of modes. This allows us to estimate the charge density for the boundary QFT$_1$ and estimate charge-charge correlation at finite density. 
\section{Scalars on NATD $ AdS_5 $}
The Non-Abelian T dual (NATD) of $ AdS_5  $ is given by the following line element\footnote{We set, $ \alpha'=1=L $.} 
\begin{align}
\label{e2.1}
&ds^2 = -\cosh^2 r dt^2 +dr^2+\frac{4 d\rho^2}{ \sinh^2 r}+\frac{4  \rho^2 \sinh^2r}{(16  \rho^2 + \sinh^4r)}(d\chi^2 + \sin^2\chi d\xi^2 ),
\end{align}
which is obtained by applying a Non-Abelian T-duality along $ SU(2) $ of $ AdS_5 $ \cite{Lozano:2017ole}.

Here, $ r $ is the radial coordinate of the global $ AdS_5 $ which shows a spacetime singularity near $ r \sim 0 $, which is an artefact of the presence of NS5 branes near the centre of NATD $ AdS_5 $. These geometries fall under the general half-BPS category of Lin, Lunin and Maldacena (LLM) \cite{Lin:2004nb}-\cite{Lin:2005nh} which are dual to irrelevant deformations of ($0+1$)d matrix model.

We introduce the following change of coordinate
\begin{align}
\cosh r = \frac{1}{\cos \gamma},
\end{align}
which yields the line element of the following form
\begin{align}
\label{e2.3}
ds^2 = \frac{1}{\cos^2 \gamma}(-dt^2 +d\gamma^2)+\frac{4 \cos^2 \gamma}{\sin^2\gamma}d\rho^2 + \frac{4 \rho^2 \sin^2 \gamma \cos^2 \gamma}{16 \rho^2 \cos^4 \gamma + \sin^4 \gamma}(d\chi^2 + \sin^2\chi d\xi^2 ).
\end{align}

Next, we introduce the radial coordinate of the Poincare patch
\begin{align}
\label{e2.4}
z= \cos \gamma 
\end{align}
where the boundary of the space-time is located near $ z= 0 $. Clearly, the $ r \sim 0 $ singularity corresponds to setting $ z\sim1 $, which is identified as the deep IR limit of the dual matrix model.

Finally, using \eqref{e2.4}, the line element \eqref{e2.3} in Poincare patch takes the following form
\begin{align}
\label{e2.5}
ds^2 =\frac{1}{z^2}(-dt^2 + f(z)dz^2) +g(z) d\rho^2 +h(\rho ,z)d\Omega_2 (\chi , \xi)
\end{align}
where we denote the above metric functions as\footnote{The radial coordinate $ z $ is bounded below and is always less than unity namely, $ 0\leq z <1 $. On the other hand, the $ \rho $ axis is in principle unbounded. However, for the purpose of the present analysis, we set an upper cut-off namely, $\rho_{max}\sim \rho_0 \gg 1$. This in turn reflects to the fact that we forget about the higher dimensional parent QFT and focus into the lower dimensional QFT in the presence of defects. The boundedness of the holographic direction imposes sort of ``completion'' for the NATD background. In what follows, in the next section \ref{sec3}, we estimate $ \frac{1}{\rho^2_0}\sim \frac{1}{L^2} $ corrections (which we call ``improvement'' terms) to the two point correlation while considering the ``coarse-grained'' approximation of the original NATD solution \cite{Lozano:2017ole}.}
\begin{align}
f(z)=\frac{1}{1-z^2}~;~g(z)= \frac{4z^2}{1-z^2}~;~h(\rho, z)=\frac{4 \rho^2 z^2 (1-z^2)}{16\rho^2 z^4 +(1-z^2)^2}.
\end{align}

We introduce the \emph{probe} scalar field action \cite{Son:2002sd}-\cite{deHaro:2000vlm}
\begin{align}
S = -\frac{1}{2}\int d^5x\sqrt{-G}(G^{MN}\partial_M \phi \partial_N \phi + m^2 \phi^2),
\end{align}
which reveals an equation of motion of the form
\begin{align}
\label{e2.8}
\frac{z}{h(\rho , z)}(1-z^2)\partial_z(z h(\rho , z)\partial_z \phi)+\frac{z}{h(\rho , z)}(1-z^2)\partial_\mu \Big(\frac{h (\rho , z)}{z(1-z^2)}G^{\mu \nu}\partial_\nu \phi \Big)-m^2 \phi =0,
\end{align}
where $ \mu $ denotes rest of the coordinates of the Poincare patch.

To proceed further, we propose a Fourier transform of the following form\footnote{Since the boundary theory corresponds to $ QFT_1 $, therefore there are no spatial momenta ($ k^i $) to deal with.}
\begin{align}
\label{e2.9}
\phi(z,\rho , t)=\int\frac{d\omega}{2\pi}e^{-i \omega t}F(z, \rho).
\end{align}

Substituting \eqref{e2.9} into \eqref{e2.8}, one finds
\begin{align}
\label{e2.10}
&z^2 (1-z^2)\partial^2_z F +\Big(\frac{4 z (1-z^2)^2}{16 \rho ^2 z^4 + (1-z^2)^2}  -z \left(z^2+1\right)\Big)\partial_z F+\frac{1-z^2}{4z^2}\partial^2_\rho F\nonumber\\
&+\frac{\left(1-z^2\right)^3}{32 \rho ^3 z^6+2 \rho z^2 \left(1-z^2\right)^2 }\partial_\rho F +(\omega^2 z^2  -m^2)F=0.
 \end{align}
 
 In the above derivation, we have switched off the dependencies of the scalar along the spatial directions associated with the two sphere parametrised by $\chi$ and $\xi$ directions. This stems from the fact that the boundary theory ``effectively'' looks like a ($0+1$) d deformed matrix model where the spatial directions do not change the ``qualitative'' behaviour of the boundary correlators. This can be seen as follows. One could in principle generalise the above analysis and propose a generalised ansatz of the form
 \begin{align}
 \label{ne11}
\phi(z,\rho , t , \xi , \chi)=\int\frac{d\omega}{2\pi}e^{-i \omega t + i k \xi}F(z, \rho , \chi)
\end{align}
where $k$ is the spatial momentum associated with the $\xi$- isometry direction.

Using \eqref{ne11}, one finds the modified equation of motion
\begin{align}
\label{ne2.12}
&z^2 (1-z^2)\partial^2_z F +\Big(\frac{4 z (1-z^2)^2}{16 \rho ^2 z^4 + (1-z^2)^2}  -z \left(z^2+1\right)\Big)\partial_z F+\frac{1-z^2}{4z^2}\partial^2_\rho F\nonumber\\
&+\frac{\left(1-z^2\right)^3}{32 \rho ^3 z^6+2 \rho z^2 \left(1-z^2\right)^2 }\partial_\rho F+\frac{1}{h(\rho , z)}\Big(\partial_\chi^2 F - \frac{k^2}{\sin^2 \chi}F\Big)+(\omega^2 z^2  -m^2)F=0.
 \end{align}

Clearly, the asymptotic expansions \eqref{e2.11}-\eqref{e2.12} still hold provided the fifth term in the parenthesis in \eqref{ne2.12} vanishes identically. This leads to the general solution in terms of the Legendre $P$ and $Q$ polynomials which take the following form
\begin{align}
&F(z, \rho , \chi)=Z(z)R(\rho)\mathcal{M}(\chi, k)\nonumber\\
&\mathcal{M}(\chi, k)=\sqrt[4]{|\sin ^2\chi |} \left(c_1 P_{-\frac{1}{2}}^{\frac{1}{2} \sqrt{4 k^2+1}}(\cos \chi )+c_2 Q_{-\frac{1}{2}}^{\frac{1}{2} \sqrt{4 k^2+1}}(\cos \chi )\right)
\end{align}
which therefore does not affect the qualitative behaviour of the boundary correlators in a sense that it only changes the overall pre-factor in \eqref{e2.21} as a result of the integration of the function $ |\mathcal{M}(\chi, k)|^2 $ over two sphere. Therefore, given the above facts, in the subsequent analysis, we choose not to excite the scalar along the directions associated with the two sphere.
\subsection{Asymptotic solutions}
Near the boundary $z \sim 0$, one could use the method of separation of variable $F=Z(z)R(\rho)$ to yield the following set of decoupled equations
\begin{align}
\label{e2.11}
&Z''+\frac{3}{z}Z'+\frac{1}{z^2}(\omega^2 z^2-m^2 +\frac{\nu^2}{4z^2} )Z=0\\
&R''+\frac{2}{\rho}R'-\nu^2 R=0
\label{e2.12}
\end{align}
where $\nu$ is the separation constant. 

As we discuss below, we have two classes of solutions depending on the choices of $ \nu $. These are categorised as, $ \nu =0 $ and $ \nu \neq 0 $ modes. When expanded near the boundary ($ z \sim 0 $), these modes exhibit different asymptotic behaviour and thereby a different holographic interpretation.

A careful analysis further reveals that like in the asymptotic UV ($ z \sim 0 $) limit, the dynamics \eqref{e2.10} is also separable in the deep IR, $ z \sim 1 $ which results into a set of equations of the form\footnote{One can show that for $ \nu=0 $ modes, the general solution corresponding to \eqref{e2.13} could be expressed as a linear superposition of Bessel's functions of the type, $Z (z) \sim c_1 I_0\left(\sqrt{2k^2 |1-z|}\right) + c_2 K_0\left(\sqrt{2k^2 |1-z|}\right)$ where $ k^2=\omega^2 -m^2 $ is a constant. Clearly, the function $ K_0(x) $ blows up in the limit, $ x \sim 0 $. One therefore needs to set, $ c_2=0 $ in order for the wave function to remain finite in the deep IR ($ z \sim 1 $).}
\begin{align}
\label{e2.13}
Z'' -\frac{1}{(1-z)}Z' +\frac{(\omega^2 -m^2)}{2(1-z)}Z = \nu^2 Z~;~R''=-4 \nu^2 R.
\end{align}
\subsubsection{$ \nu =0$ mode}
We first figure out the asymptotic solutions of \eqref{e2.11} corresponding to $ \nu =0 $ modes. We propose an asymptotic solution of the form, $ Z_0 \sim z^\beta $ where $\beta$ is some real exponent. Substituting into \eqref{e2.11}, and keeping the leading order terms in the expansion near the boundary one finds
\begin{align}
\beta = -1\pm n~;~ n= \sqrt{m^2+1}.
\end{align}

Therefore, the general solution could be expressed as
\begin{align}
\label{e2.14}
Z^{(\omega)}_0 (z)=\mathcal{A}(\omega)z^{-(n+1)}+\mathcal{B}(\omega)z^{n-1}.
\end{align}

Clearly, for $ \beta $ to be real, one must satisfy $ m^2 \geq  -1$ which is the analogue of Breitenlohner-Freedman(BF) bound for NATD $ AdS_{4+1} $. Notice that, for $ n ( \geq 0 ) $ to be a real positive number, the first term on the r.h.s. of \eqref{e2.14} is always dominant near the boundary, $ z \sim \epsilon \sim 0 $ as compared to the second term. Therefore, the complete solution could be expressed as
\begin{align}
&\phi_{\nu =0}(z \sim \epsilon , t, \rho)=Z_{0}(\epsilon , t)R_{0}( \rho)\\
&R_{0}(\rho)=c_2-\frac{c_1}{ \rho}.
\label{e2.16}
\end{align}

The asymptotic radial mode is given by
\begin{align}
Z_{0}(\epsilon, t)=\mathcal{A}(t)\epsilon^{-(n+1)}+\mathcal{B}(t)\epsilon^{n -1}
\end{align}
where the first term is clearly divergent as we shift the UV cut-off to zero, $\epsilon\rightarrow 0$. 

In order to establish the holographic dictionary \cite{Ramallo:2013bua}-\cite{Natsuume:2014sfa}, we rescale the boundary mode
\begin{align}
\label{e2.19}
\hat{Z}_0 (\epsilon,t)=\lim_{\epsilon \rightarrow 0}\epsilon Z_{0}(\epsilon, t)=\mathcal{A}(t)\epsilon^{-n}+\mathcal{B}(t)\epsilon^{n}
\end{align}
where we define the boundary source and the dual operator in terms of this rescaled mode.

For example, we introduce the source for the boundary matrix model as
\begin{align}
\label{e2.20}
\mathcal{A}(t)=J(t)= \lim_{\epsilon \rightarrow 0}\epsilon^{n}\hat{Z}_{0}(\epsilon, t)
\end{align}
where we identify,  $\mathcal{B}(t) \sim \mathcal{O} (t)  $ as the corresponding dual operator. This can be seen explicitly by considering a boundary action for the dual ($ 0+1 $)d matrix model
\begin{align}
\label{e2.21}
&\hat{S}_{bdy}=16 \pi \int_0^{\rho_0} \rho^2 R^2_0(\rho) d\rho \int dt \hat{Z}_{0}(\epsilon, t)\mathcal{O}(\epsilon, t)\nonumber\\
&=16 \pi \int_0^{\rho_0} \rho^2 R^2_0(\rho) d\rho\int dtJ(t)\mathcal{O}(t)
\end{align}
where $ \hat{S}_{bdy}$ is the true boundary action that represents the dual matrix model. 

In the above, we use \eqref{e2.19} and introduce the dual operator
\begin{align}
\mathcal{O}(\epsilon, t)= \lim_{\epsilon \rightarrow 0}\epsilon^{n+1} \mathcal{O}(t).
\end{align}

We scale the boundary action \eqref{e2.21} by the overall (volume and $ \rho $- integral) pre-factor and interpret \eqref{e2.21} as an deformation of the boundary ($ 0+1 $)d matrix model of the form
\begin{align}
\Delta S_{QFT}=\lambda\int dt J(t)  \mathcal{O}(\epsilon , t)
\end{align}
where $ \lambda $ is the associated coupling constant with negative mass dimension, $[\lambda]=-n$. 

We have two distinct cases here namely - (i) for $ m^2=-1$, the mass dimension becomes $ n =0 $, which corresponds to turning on a \emph{marginal} operator and (ii) for $ m^2 > -1$, we find $ n > 0 $ which corresponds to turning on an \emph{irrelevant} operator for the dual ($ 0+1 $)d  matrix model. Clearly, as one can see, in either of these situations the scalar deformation cannot be a \emph{relevant} one. The purpose of this paper is to discuss the consequences of each of the above deformations on the two point correlation in the dual matrix model at strong coupling. 
\subsubsection{$ \nu \neq 0$ mode}
For $ \nu \neq 0 $, the general solution of \eqref{e2.11} could be expressed in terms of Bessel's functions
\begin{align}
\label{e2.13}
Z^{(\omega)}_\nu (z)=\frac{\nu}{4 z} \Big(a_\nu(\omega) \Gamma \left(n+1\right) J_{n}\left(\frac{\nu }{2 z}\right)+b_\nu(\omega) \Gamma \left(|1-n|\right) J_{-n}\left(\frac{\nu }{2 z}\right)\Big).
\end{align}

Notice that, the argument of the Bessel function diverges as we approach the boundary $ z \sim 0 $. A careful expansion near the boundary further reveals
\begin{align}
&Z^{(\omega)}_\nu (z \sim 0)\simeq \frac{\sqrt{\nu}}{\sqrt{\pi z}}\Big[a_\nu(\omega)\Gamma (n+1)\sin \left(\frac{ \pi }{4} (1-2n)+\frac{\nu}{2z}\right)\nonumber\\
&+b_\nu(\omega)\Gamma (|1-n|)\sin \left(\frac{ \pi }{4} (1+2n)+\frac{\nu}{2z}\right)\Big].
\end{align}

The complete solution turns out to be
\begin{align}
\phi_{\nu}(z \sim 0,\rho)=Z^{(\omega)}_\nu (z \sim 0)R_\nu (\rho)
\end{align}
where we denote the solution along the $ \rho $ - axis as
\begin{align}
\label{e2.27}
R_\nu (\rho) =\frac{2 c_1 \nu  e^{-\nu  \rho }+c_2 e^{\nu  \rho }}{2 \nu  \rho }.
\end{align}
\subsection{Green's function}
The goal of this section is to compute two point correlations between operators $ \mathcal{O} (t) $ in the dual matrix model by studying scalar field dynamics in the NATD background \eqref{e2.5}. The standard technique employed would be the celebrated holographic correspondence between the asymptotic radial modes ($ \sim Z_0 $) and the dual operators ($ \mathcal{O} (t) $) in the matrix model.  

As we show below, we have different asymptotics corresponding $\nu =0$ and $\nu \neq 0$ modes. While the holographic interpretation is clear for $\nu =0$ modes, it is not so obvious for $\nu \neq 0$ modes. In other words, correlation functions are meaningful only for $\nu =0$ modes.
\subsubsection{$  \nu =0$ mode}
The on-shell boundary action \cite{Son:2002sd}-\cite{deHaro:2000vlm} could be expressed as\footnote{The surface term corresponding to the $ \rho $- integral produces a divergent contribution that goes like, $ S_\rho \sim \rho^2_0 $ in the bulk ($ 0<z<1 $) where $ \rho_0 $ is the upper bound along $ \rho $- direction which is in principle unbounded for the NATD example. However, when explored close to the boundary ($ z \sim \epsilon $), one finds divergent contribution of the form, $ S_\rho |_{z = \epsilon}\sim \epsilon^{-2n} $ which has a similar pole structure as that of \eqref{e2.35} and therefore can be tamed by introducing an appropriate counter term near the boundary.}
\begin{align}
\label{e2.28}
\hat{S}_{bdy} =16 \pi \int_0^{\rho_0}d\rho\rho^2  R^2_0 (\rho) \int dt \hat{\Pi} (\epsilon , t)\hat{Z}_0 ( \epsilon , t)\Big |_{z=\epsilon}
\end{align}
where we introduce the rescaled canonical momentum as
\begin{align}
\hat{\Pi} (\epsilon , t)=\lim_{\epsilon \rightarrow 0}\epsilon^2  \partial_z Z_0|_{z=\epsilon}=- (n+1) \mathcal{A}(t)\epsilon^{-n}+(n -1)\mathcal{B}(t)\epsilon^{n}.
\end{align}

Next, we use the Fourier transform to map these solutions into momentum space
\begin{align}
\label{e2.30}
&\hat{\Pi}_{(\omega)}(\epsilon)=\int \frac{d\omega}{2\pi}e^{-i \omega t}\hat{\Pi} (\epsilon , t)=- (n+1) \mathcal{A}(\omega)\epsilon^{-n }+(n-1)\mathcal{B}(\omega)\epsilon^{n}\\
&\hat{Z}^{(\omega)}_0 (\epsilon)=\int \frac{d\omega}{2\pi}e^{-i \omega t}\hat{Z}_0 (\epsilon , t)=\mathcal{A}(\omega)\epsilon^{-n}+\mathcal{B}(\omega)\epsilon^{n}.
\label{e2.31}
\end{align}

Using \eqref{e2.30}-\eqref{e2.31}, the on-shell boundary action \eqref{e2.28} finally takes the form
\begin{align}
\label{e2.34}
& \hat{S}_{bdy}=16 \pi \int_0^{\rho_0}\rho^2 R^2_0(\rho) d\rho \int \frac{d\omega}{2 \pi}\hat{\Pi}_{(-\omega)}(\epsilon) \hat{Z}^{(\omega)}_0 (\epsilon)\nonumber\\
&=16 \pi \Lambda_0 \int\frac{d \omega}{2 \pi} \Big ( - (n+1) \mathcal{A}(-\omega)\mathcal{A}(\omega) \epsilon^{-2n}-(n+1) \mathcal{A}(-\omega)\mathcal{B}(\omega)\nonumber\\
&+ (n-1)\mathcal{A}(\omega)\mathcal{B}(-\omega) \Big)+\mathcal{O}(\epsilon^{2n})
\end{align}
where the constant pre-factor, $ \Lambda_0 $ is an artefact of the $ \rho $- integration in \eqref{e2.34}.

Clearly, the leading term in \eqref{e2.34} is divergent in the boundary limit, $ \epsilon \rightarrow 0 $. To get rid of this divergence, one therefore needs to add a suitable counter term 
\begin{align}
\label{e2.35}
S^{(ct)}=\frac{\epsilon (n+1)}{2}\int d\chi d \xi d\rho dt \sqrt{-G}\phi^2_{\nu =0}\Big |_{z=\epsilon}=16 \pi \Lambda_0 (n+1)  \int dt \hat{Z}_0^2(\epsilon , t).
\end{align}

Using the Fourier transform \eqref{e2.31}, one could further expand \eqref{e2.35} to yield
\begin{align}
\label{e2.36}
&S^{(ct)}=16 \pi \Lambda_0(n+1) \int \frac{d \omega}{2 \pi}\Big(\mathcal{A}(-\omega)\mathcal{A}(\omega) \epsilon^{-2n}+ \mathcal{A}(\omega)\mathcal{B}(-\omega)\nonumber\\&+  \mathcal{A}(-\omega)\mathcal{B}(\omega)\Big)+\mathcal{O}(\epsilon^{2n}).
\end{align}

Using \eqref{e2.36}, the renormalised boundary action finally turns out to be
\begin{align}
S^{(ren)}_{bdy}=32 \pi n \Lambda_0\int \frac{d \omega}{2 \pi}\mathcal{A}(\omega)\mathcal{B}(-\omega)+\mathcal{O}(\epsilon^{2n}).
\end{align}

The two point function is given by
\begin{align}
\label{e2.37}
\mathcal{G}(\omega) =32 \pi n\Lambda_0 \frac{\mathcal{B}(-\omega)}{\mathcal{A}(\omega)}.
\end{align}

The evaluation of the ratio $ \frac{\mathcal{B}}{\mathcal{A}} $ is in general a non-trivial task for the present background. However, we estimate this ratio under certain specific assumptions\footnote{We construct the solution in a region where, $z \sim 1/\rho_0$. This is essentially the limit where one could ``decouple" the zero modes from holographic ($\rho$) coordinate (that arises as an artefact of NATD which in turn introduces a ``defect'' in the ($0+1$)d QFT) by considering the (holographic) RG direction ($z$) to be small enough.}, where we construct the radial mode equation away from the boundary limit, $ z \sim 0 $ and write down its general solution
\begin{align}
\label{ee2.38}
&Z_0(z) = z^{-1-n}\Big(\mathcal{A}(\omega)\, _2F_1\left(\frac{1}{2} (-n-\omega -1),\frac{1}{2} (-n+\omega -1);1-n;z^2\right) \nonumber\\
&+ (-1)^n \mathcal{B}(\omega) z^{2n}\, _2F_1\left(\frac{1}{2} (n-\omega -1),\frac{1}{2} (n+\omega -1);n+1;z^2\right)\Big)
\end{align}
in terms of Hyper-geometric functions \cite{book}.

When expanded close to the boundary, \eqref{ee2.38} produces correct expansion as in \eqref{e2.14}. On the other hand, an expansion near $z \sim 1$ reveals
\begin{align}
\label{e2.39}
&Z_0(z \sim 1) = \frac{\mathcal{A}(\omega) \Gamma (|1-n|)}{\Gamma \left(\frac{1}{2} (-n-\omega +3)\right) \Gamma \left(\frac{1}{2} (-n+\omega +3)\right)}\nonumber\\&+\frac{ (-1)^n \mathcal{B}(\omega)\Gamma (n+1)}{\Gamma \left(\frac{1}{2} (n-\omega +3)\right) \Gamma \left(\frac{1}{2} (n+\omega +3)\right)}+\mathcal{O}(1-z).
\end{align}

By imposing the fact that the solution \eqref{e2.39} vanishes smoothly in the IR, one finds the ratio $\frac{\mathcal{B}}{\mathcal{A}} $. Considering a small frequency expansion, this finally yields the two point correlation 
\begin{align}
\label{e2.40}
&\mathcal{G}(\omega \sim 0) \sim 32 \pi n \Lambda_0 \Big(\frac{\Gamma (|1-n|) \Gamma \left(\frac{n+3}{2}\right)^2}{\Gamma \left(\frac{3}{2}-\frac{n}{2}\right)^2 \Gamma (n+1)}\nonumber\\
&+\frac{\omega ^2 \Gamma (|1-n|) \Gamma \left(\frac{n+3}{2}\right)^2 \left(\psi ^{(1)}\left(\frac{n+3}{2}\right)-\psi ^{(1)}\left(\frac{3}{2}-\frac{n}{2}\right)\right)}{4 \Gamma \left(\frac{3}{2}-\frac{n}{2}\right)^2 \Gamma (n+1)} \Big)+\mathcal{O}(\omega^4)
\end{align}
where $\psi^{(p)}(z)$ denotes polygamma function \cite{book} of order $ p $.

A closer look reveals that the correlation function \eqref{e2.40} vanishes for marginal operators, $n =0$ and is ill-defined for a discrete set of values $n=1,3, \cdots$. On the other hand, we have a set of well defined scalar correlators for $n= 2, 4, \cdots$. For example, considering $n=2$, the corresponding two point function \eqref{e2.40} turns out to be
\begin{align}
\mathcal{G}(\omega \sim 0) \sim 64 \pi \Lambda_0 \Big(\frac{9}{32} -\frac{5}{16} \omega^2 + \cdots \Big).
\end{align}

On the other hand, at high frequencies, the Green's function behaves as
\begin{align}
\mathcal{G}(\omega \gg 1) \sim \omega ^n.
\end{align}

Clearly, at high frequencies, one could distinguish between different scalar operators by looking at the corresponding two point correlations which are different for distinct choices of the parameter $n$. This indicates to the fact that the operator properties are more distinguishable at higher frequencies, while they exhibit identical spectrum in the domain of low frequencies.
\subsubsection{$ \nu \neq 0$ mode}
The wave function \eqref{e2.9} near the UV asymptotic ($ z \sim \epsilon \sim 0 $) turns out to be
\begin{align}
\label{e2.46}
&\phi_\nu (z\sim \epsilon, \rho, t)=\int\frac{d\omega}{(2\pi)}e^{-i \omega t}Z^{(\omega)}_\nu (\epsilon)R_\nu (\rho)\nonumber\\
&= \frac{\sqrt{\nu}}{\sqrt{\pi \epsilon}}\Big [a_\nu(t)\Gamma (n+1)\sin \left(\frac{ \pi }{4} (1-2n)+\frac{\nu}{2\epsilon}\right)\nonumber\\
&+b_\nu(t)\Gamma (1-n)\sin \left(\frac{ \pi }{4} (1+2n)+\frac{\nu}{2\epsilon}\right) \Big]R_\nu (\rho).
\end{align}

Using \eqref{e2.46}, the (on-shell) boundary action turns out to be\footnote{The surface term due to $\rho$- integral could be expressed as $S_\rho \sim \rho_0^2$ where the proportionality constant depends on the value of the $z$ integration which ranges between $0$ to $1$. However, a careful analysis near the UV region $z \sim \epsilon \sim 0$ yields a divergent contribution of the form, $S_\rho \sim e^{2 \nu \rho_0}\int \frac{d\epsilon}{\epsilon} |Z_\nu^{(\omega)}(\epsilon)|^2$. Clearly, the structure of the UV divergence is different from what we have observed in the case of $\nu =0$ modes. The radial mode $Z_\nu^{(\omega)}(\epsilon)$ contains harmonic functions that are ill-defined in the limit, $\epsilon \rightarrow 0$ which makes it even harder to understand the precise meaning of such divergences near the boundary.}
\begin{align}
\label{e2.49}
S_{bdy}=\frac{1}{2}V_i\int dt d\rho \sqrt{-G}G^{zz}\partial_z \phi_\nu \phi_\nu \Big |_{z \sim \epsilon }
\end{align}
where $ V_i $ is the volume of the two sphere.

Like before, we can express \eqref{e2.49} in frequency space using the inverse Fourier transform. To do that, we first note-down the following expansions
\begin{align}
\label{e2.50}
&\Pi_\nu^{(\omega)}(z \sim \epsilon)=\frac{-4 \rho^2 \nu^{3/2} \sqrt{\epsilon}}{\sqrt{\pi}}\Big[a_\nu(\omega)\Gamma (n+1)\cos \left(\frac{ \pi }{4} (1-2n)+\frac{\nu}{2\epsilon}\right)\nonumber\\
&+b_\nu(\omega)\Gamma (1-n)\cos \left(\frac{ \pi }{4} (1+2n)+\frac{\nu}{2\epsilon}\right)  \Big]=\frac{-4 \rho^2 \nu^{3/2}\sqrt{\epsilon} }{\sqrt{\pi}}\tilde{\Pi}^{(\omega)}_\nu (\epsilon).
\end{align}

Using \eqref{e2.50}, the boundary action \eqref{e2.49} finally turns out to be
\begin{align}
\label{e2.51}
S_{bdy}\Big |_{z \sim \epsilon}=-\frac{4\nu^2}{\pi}V_i \int_0^{\rho_0}d\rho \rho^2 R^2_\nu (\rho)\int \frac{d \omega}{2 \pi}\tilde{\Pi}_\nu^{(-\omega)}(\epsilon) Z_\nu^{(\omega)}(\epsilon)\Big |_{z \sim \epsilon}.
\end{align}

The $ \rho $- integration reveals a constant pre-factor
\begin{align}
 \int_0^{\rho_0}d\rho \rho^2 R^2_\nu (\rho)=\frac{1}{8 \nu ^3}(4 c_1 \nu ^2 \left(c_1 \left(-e^{-2 \nu  \rho_0}\right)+2 c_2 \rho_0+c_1\right)+c_2{}^2 \left(e^{2 \nu  \rho_0}-1\right)).
 \end{align}
 
Finally, a straightforward evaluation of the second integral in \eqref{e2.51} yields
\begin{align}
&\int \frac{d \omega}{2 \pi}\tilde{\Pi}_\nu^{(-\omega)}(\epsilon) Z_\nu^{(\omega)}(\epsilon)\Big |_{z \sim \epsilon}=\int \frac{d \omega}{4 \pi} \Big(a_\nu (- \omega)a_\nu (\omega)\Gamma(1+n)^2 \cos(n \pi - \frac{\nu}{\epsilon})\nonumber\\&+(a_\nu (- \omega)b_\nu (\omega)(\sin (\pi  n)+\cos \left(\frac{\nu }{\epsilon}\right))+a_\nu ( \omega)b_\nu (-\omega)(\cos \left(\frac{\nu }{\epsilon}\right)-\sin (\pi  n)))\Gamma(1+n)\Gamma(1-n) \nonumber\\&+ b_\nu (- \omega)b_\nu (\omega)\Gamma(1-n)^2 \cos(n \pi + \frac{\nu}{\epsilon})\Big)
\end{align}
which is a rapidly oscillating function in the boundary limit, $ \epsilon \sim 0 $. 

In order to carry out a holographic computation, one has to find a way to renormalise the on-shell boundary action \eqref{e2.49}. However, this turns out to be a nontrivial task as the trigonometric functions become ill-defined in the limit, $\epsilon \rightarrow 0$. In other words, it is difficult to make sense of the boundary action \eqref{e2.51} for $ \nu \neq 0 $ modes, which makes it even harder to find holographic interpretations for such excitations. As we elaborate, these modes correspond to excitations of a ``defect'' ($0+1$)d QFT where the defect is sourced due to the holographic ($\rho$) direction. A detailed understanding of these modes we postpone until section \ref{sec3.3}.
\section{Remarks about coarse-grained solution}
\label{sec3}
As conjectured by authors in \cite{Lozano:2017ole}, the NATD background considered in \eqref{e2.1} corresponds to an irrelevant deformation of $\mathcal{N}=2$ supersymmetric matrix model. Therefore, in order for the NATD solution to represent a PWMM which characterises D0 brane plus some \emph{relevant} deformation, the authors in \cite{Lozano:2017ole} proposed a ``coarse-grained'' approximation in the Lin-Maldacena description of conducting plates \cite{Lin:2005nh}. They introduce a potential function of the form\footnote{Here, $ \alpha = \frac{L^4}{16\alpha'^2} $ is a dimensionless parameter of the theory \cite{Lozano:2017ole}.} \cite{Lozano:2017ole}
\begin{align}
\label{e3.1}
&V(\sigma , \eta )=(\eta \sigma^2 -\frac{2}{3}\eta^3)-2 \alpha  \eta  \log \sigma + \alpha  \sqrt{(L-\eta )^2+\sigma ^2}-\alpha  \sqrt{(\eta +L)^2+\sigma ^2}\nonumber\\
&+\alpha  \eta  \log \left(\left(\sqrt{(L-\eta )^2+\sigma ^2}+(L-\eta )\right) \left(\sqrt{(\eta +L)^2+\sigma ^2}+(\eta +L)\right)\right)
\end{align}
where the coordinate $ \eta = 2\rho \in [0, L] $ is finite to start with. On the other hand, here $ \sigma = \cosh r $. 

This is essentially cutting the holographic axis at a finite but large L, such that one cannot reach the parent QFT living in some higher dimensions. In other words, we focus on the QFT living in ($ 0+1 $)d and do not care about the parent QFT from which it is derived. Mapping from a higher dimensional QFT to a lower ($0+1$) dimensional QFT is a trademark signature of the NATD solution we discuss in this paper. Therefore, the above truncation ``effectively'' acts like a relevant deformation, where we disregard the degrees of freedom beyond scale, $\rho_0 \sim L$.

The potential function \eqref{e3.1} satisfies Laplace's equation 
\begin{align}
\ddot{V}+\sigma^2 V'' =0
\end{align}
where $ \dot{V}=\sigma \partial_\sigma V $ and $ V'=\partial_\eta V $ and so on.

Considering a large $ L \rightarrow \infty $ limit, yields an expansion of the form
\begin{align}
V(\sigma , \eta)=2V_{back}+\alpha  \eta  \left(\log \left(4 L^2\right)-2 \right)+\frac{\alpha}{L^2}\Big(\frac{3 \eta  \sigma ^2}{2}-\eta ^3\Big)+\mathcal{O}(L^{-4})
\end{align}
where we identify
\begin{align}
V_{back}(\sigma , \eta)=\frac{\eta  \sigma ^2}{2}-\frac{ \eta ^3}{3}- \alpha \eta \log \sigma
\end{align}
as the potential function of the constant flux background \cite{Lin:2005nh} which corresponds to smeared D0 branes at asymptotic infinity, $ \sigma \rightarrow \infty $ and $ \eta = $ fixed.

Given the potential function \eqref{e3.1}, one could express the 10d metric as\footnote{In what follows, in the subsequent analysis, we switch off $ S^5 $ and focus on the scalar field dynamics in the ``cured'' NATD of $ AdS_5 $ where we retain ourselves upto leading order in the $ \frac{1}{L^2} $ corrections.}
\begin{align}
\label{e3.5}
ds^2_{10}=-f_1(\sigma , \eta) dt^2 +f_2 (\sigma , \eta) (d\sigma^2 + d\eta^2)+f_3 (\sigma , \eta) d\Omega_2 (\chi , \xi)+f_4 (\sigma , \eta)  d\Omega_5 
\end{align}
where the individual metric functions could be expressed as
\begin{align}
&f_1(\sigma , \eta)=\frac{4 \ddot{V}}{\sqrt{V''(2\dot{V}-\ddot{V})}}~;~f_2(\sigma , \eta)=\frac{8 \ddot{V}}{\dot{V}}\frac{1}{f_1}\\
&f_3(\sigma , \eta)=-\frac{8 \dot{V}\ddot{V}}{\Delta}\frac{1}{f_1}~;~f_4(\sigma , \eta)=-\frac{16 \ddot{V}}{V''}\frac{1}{f_1}\\
&\Delta =(\ddot{V}-2\dot{V})V''  -(\dot{V}')^2.
\end{align}

One could further simplify these metric coefficients upto leading order in the $ 1/L^2 $ expansion\footnote{Translating into the language of the previous analysis, this simply corresponds to finding $ 1/\rho^2_0 \sim 1/L^2 $ corrections to the results obtained previously in the NATD example.}
\begin{align}
&f_1(\sigma , \eta)=\sigma^2\Big( 4+\frac{3}{L^2} \Big)~;~f_2(\sigma , \eta)=\frac{4}{(\sigma^2 -1)}\Big( 1-\frac{3k(\sigma)}{4L^2} \Big)~;~ k(\sigma)=\frac{\sigma^2 +1}{\sigma^2 -1}\\
&f_3(\sigma , \eta)=\frac{4 \eta^2  \left(\sigma ^2 -1\right)}{4 \eta ^2+\left(\sigma ^2 -1\right)^2}-\frac{3}{L^2}s(\sigma , \eta)~;~s(\sigma , \eta)=\frac{ \eta^2 \left(\sigma ^2+1\right) \left(\left(\sigma ^2-1\right)^2-4 \eta ^2\right)}{ \left(4 \eta ^2+\left(\sigma ^2-1\right)^2\right)^2}
\end{align}
where we set, $ \alpha =1 $ without any loss of generality.

To proceed further, we use the following change of coordinate
\begin{align}
\sigma = \frac{1}{z}
\end{align}
and re-express the 5d background \eqref{e3.5} in Poincare coordinate
\begin{align}
\label{e3.12}
ds_5^2 = -f_1(z) dt^2 +f_2 (z) dz^2 + z^4 f_2(z)d\eta^2+f_3 (z , \eta)d\Omega_2 (\chi , \xi).
\end{align}

The metric functions could be easily read off
\begin{align}
\label{e3.13}
&f_1(z) =\frac{a}{z^2}~;~a=\Big( 4+\frac{3}{L^2} \Big)\\
&f_2(z) =\frac{4}{z^2(1-z^2)}\Big( 1-\frac{3k(z)}{4L^2} \Big)~;~k(z)=\frac{1+z^2}{1-z^2}\\
&f_3 (z , \eta)=\frac{4 \eta^2 z^2 \left(1-z^2\right)}{4 \eta ^2 z^4+\left(1 - z^2\right)^2}-\frac{3}{L^2}s(z , \eta)
\label{e3.15}
\end{align}
which smoothly reduces to \eqref{e2.5} in the strict $ L \rightarrow \infty $ limit (modulo an overall pre-factor of $ 4 $), along with a proper identification between $ \eta $ and $ \rho $ coordinates as mentioned before. Notice that, if we switch-off the holographic $ \eta $- axis and do not excite the two sphere ($S^2$), the background \eqref{e3.12} eventually flows into an $ AdS_2 $ which is dual to a conformal quantum mechanics in ($ 0+1 $)d.
\subsection{Scalar dynamics}
We choose to work in the domain of finite $\eta$, where $ \eta $ lives in an interval $ [0,L] $ with $ \eta_{max}=L (=2\rho_0)$ \cite{Lozano:2017ole}. Towards the end of our calculation, we take the boundary limit, $z \sim 0$ and integrate out the $\eta$ direction which results in a dual description of a QFT living in ($0+1$)d.

To start with, the equation of motion for the scalar field looks like
\begin{align}
\frac{1}{f_2}\Big(\partial^2_z \phi +\frac{1}{z}\Big(1+\frac{z\partial_z f_3}{f_3}  \Big)\partial_z \phi \Big)+\frac{1}{z^4 f_2}\Big( \partial^2_\eta \phi +\frac{\partial_\eta f_3}{f_3} \partial_\eta \phi \Big)-\frac{z^2}{a}\partial^2_t \phi - m^2 \phi =0.
\end{align}

By expanding the metric functions $ f_i(z, \eta) $ upto $ 1/L^2 $ and substituting\footnote{Like before, we switch-off the angular ($\chi$ and $\xi$) dependencies of the scalar since this will not affect the qualitative behaviour of the boundary matrix model correlators.}
\begin{align}
\phi ( z, \eta , t)=\int \frac{d \omega}{2 \pi}e^{-i \omega t}F(z , \eta)
\end{align}
one finds the $ 1/L^2 $ corrected equation
\begin{align}
&\Big(\frac{z^2}{4} \left(1-z^2\right)+\frac{3z^2 \left(1+z^2\right)}{16 L^2}\Big) \partial^2_z F +\Big( \frac{z \left(z^2-1\right)^2}{4 \eta ^2 z^4+\left(z^2-1\right)^2}-\frac{1}{4} z \left(z^2+1\right)\nonumber\\ &+ \frac{3 z \left(z^2-1\right) \left(16 \eta ^4 z^8+\left(z^2-1\right)^3 \left(z^2+3\right)+8 \eta ^2 \left(z^2-5\right) \left(z^2+1\right) z^4\right)}{16 L^2 \left(4 \eta ^2 z^4+\left(z^2-1\right)^2\right)^2}  \Big)\partial_z F \nonumber\\
&+\Big( \frac{1-z^2}{4 z^2}+\frac{3 \left(z^2+1\right)}{16 L^2 z^2} \Big)\partial^2_\eta F +\Big(\frac{\left(1-z^2\right)^3}{8 \eta ^3 z^6+2 \eta  \left(z^2-1\right)^2 z^2} \nonumber\\&+\frac{3 \left(z^2-1\right)^2 \left(z^2+1\right) \left(12 \eta ^2 z^4+\left(z^2-1\right)^2\right)}{8 \eta  L^2 \left(\left(4 \eta ^2+1\right) z^5-2 z^3+z\right)^2} \Big)\partial_\eta F+\Big(\frac{\omega^2 z^2}{4}(1-\frac{3}{4L^2})-m^2 \Big)F=0.
\end{align}

Like before, considering an expansion near the boundary ($ z \sim 0 $), one is able to carry out a separation of variables $ F(z , \eta) =Z(z)R(\eta)$ which leads to the following set of equations
\begin{align}
\label{e3.19}
&\frac{a}{4}Z'' + \frac{3 a}{4 z} Z'+\frac{1}{z^2}\Big(\frac{4\omega^2 z^2}{a}-4m^2 +\frac{\nu^2}{4z^2} \Big)Z=0\\
&R'' +\frac{2}{\eta}R'-\frac{\nu^2}{a}R=0
\end{align}
where $ a $ is defined in \eqref{e3.13}.
\subsection{Improved correlations}
We first focus on $ \nu=0 $ modes. These modes allows an asymptotic expansion of the form, $ Z_0 \sim z^\beta $ which upon substitution into \eqref{e3.19} yields
\begin{align}
\beta =-1 \pm \hat{n} ~;~ \hat{n}=\sqrt{1+ \hat{m}^2}
\end{align}
where we defined the re-scaled mass of the scalar
\begin{align}
\hat{m}^2 \simeq 4m^2 \Big( 1-\frac{3}{4L^2}\Big).
\end{align}

Qualitatively, the discussion for the zero modes proceeds in an identical fashion except for the fact that the mass of the scalar is now re-scaled. In other words, the general structure \eqref{e2.37} remains the same while the ratio $ \frac{\mathcal{B}}{\mathcal{A}} $ could be estimated by constructing the radial solution away ($ z \sim \frac{1}{L} $) from the boundary and considering its consistency while approaching the IR ($ z \sim 1 $). 

The resulting procedure yields a solution in terms of Hyper-geometric functions \cite{book} 
\begin{align}
&Z_0 (z) \sim \Big( \frac{2}{L}\Big)^{-(1+n)}(3+4L^2)^{\frac{1-n}{2}}z^{-1-n}\Big[(-1)^{\frac{1+n}{2}} (3+ 4L^2)^n \mathcal{A}(\omega)\nonumber\\
&\, _2F_1\left(\frac{1}{4} \left(-2n-\frac{\sqrt{4 L^2-3} \omega }{L}-2\right),\frac{1}{4} \left(-2n+\frac{\sqrt{4 L^2-3} \omega }{L}-2\right);1-n;\frac{4 L^2 z^2}{4 L^2+3}\right)+(-1)^{\frac{1-n}{2}}\nonumber\\
&(2L)^{2n}z^{2n} \mathcal{B}(\omega)\, _2F_1\left(\frac{1}{4} \left(2n-\frac{\sqrt{4 L^2-3} \omega }{L}-2\right),\frac{1}{4} \left(2n+\frac{\sqrt{4 L^2-3} \omega }{L}-2\right);n+1;\frac{4 L^2 z^2}{4 L^2+3}\right)\Big]
\end{align}
which yields correct expansion near the UV asymptotic, $ z \sim 0 $.

Following our previous procedure, the smoothness of the radial solution near $ z \sim 1 $ yields
\begin{align}
&\frac{\mathcal{B}}{\mathcal{A}}= \text{const.}+\frac{ 4^{-n-1} \omega ^2  a^n \Gamma (|1-n|) \Gamma \left(\frac{n+3}{2}\right)^2}{\Gamma \left(\frac{3}{2}-\frac{n}{2}\right)^2 \Gamma (n+1)}\nonumber\\
&\times  \left(\psi ^{(1)}\left(\frac{3}{2}-\frac{n}{2}\right)-\psi ^{(1)}\left(\frac{n+3}{2}\right)\right)+\mathcal{O}(\omega^4).
\end{align}

A direct comparison with \eqref{e2.40} reveals that the correlation function is corrected by a factor $ a^n \simeq 4^n (1+\frac{3n}{4L^2}) $ at leading order in the $ 1/L^2 $ expansion, where $ \frac{3n}{4L^2} $ is the improvement at leading order. Clearly, in the strict $ L \rightarrow \infty $ one recovers the $ \omega^2 $ correction of \eqref{e2.40}.

Finally, for $ \nu \neq 0 $ modes, one finds the asymptotic behaviour as
\begin{align}
Z^{(\omega)}_\nu (z \sim 0)=\frac{\nu (4a)^n}{z(2\sqrt{a})^{2n+1}}\Big(a_\nu (\omega)  \Gamma (1+n)J_{n}\left(\frac{\nu }{\sqrt{a} z}\right) +b_\nu (\omega) \Gamma (|1-n|)J_{-n}\left(\frac{\nu }{\sqrt{a} z}\right)\Big)
\end{align}
which leads to rapidly oscillating harmonic functions and thereby a ``non-renormalisable'' on-shell action at the boundary.
\subsection{A qualitative discussion and comparison of results}
\label{sec3.3}
We now present a ``bulk argument'' and draw a comparative analysis between the results that are obtained so far. This further pave the way for a similar understanding for vector modes which we discuss in the next section. A closer look at \eqref{e2.5} or \eqref{e3.12} reveals that the geometry approaches a ``nearly'' $ AdS_2 $ asymptotic in the boundary ($z \sim 0$) limit. Therefore, the boundary matrix model could be recast as an ``approximate'' conformal quantum mechanics in ($ 0+1 $)d which include deformations that are controlled by the UV cut-off ($\epsilon$) of the theory. In the presence of the holographic ($ \rho $ or $ \eta $) direction, the dual matrix model behaves like a ``defect'' QFT$_1 $ derived from some higher dimensional parent QFT. There have been speculations about this parent QFT, however for the NATD $ AdS_5 $, the parent theory could be thought of as the $ \mathcal{N}=4 $ SYM living in four space time dimensions. The (unbounded) $ \rho $ direction therefore imposes a defect on the lower dimensional theory and carries the information about this parent theory and somehow forbids us to access the full information. Therefore, in order to carry out a meaningful calculation, one needs to put some cut-off along the holographic $ \rho $ (or $ \eta $) axis, where we forget about this infinite possible information and focus only on ($0+1$)d defect QFT.

When we discuss $\nu =0$ modes, a closer look at the structure of the underlying equations of motion \eqref{e2.11} reveals that they are essentially ignorant about the holographic ($\rho$ or $\eta$) direction. In other words, these modes do not explicitly ``see'' the defect (at the level of their dynamics) as is introduced by the holographic $\rho$ (or the $\eta$)  direction as an artefact of the NATD. This eventually turns out to be the reason that we are able to construct scalar operators and study their correlation functions that are pertinent to a ($ 0+1 $)d defect QFT. 

On the other hand, for $\nu \neq 0$ modes, one essentially excites the $ \rho $ (or the $ \eta $) direction\footnote{For more clarity, see the discussion below \eqref{e4.5}. Identical arguments hold for scalar modes as well.} which enters into the equations of motion. In other words, these modes carry the memory of ``defect'' that enters into the ($0+1$) dimensional QFT.  As a natural consequence of this, unlike the zero modes, we loose the information about the scalar correlators of the theory. This is the reason we cannot interpret these modes in the standard framework of the holographic correspondence.
\section{Vectors on NATD $ AdS_5 $}
We now extend our analysis for vector perturbations on NATD $ AdS_5 $. The starting point is the Maxwell's action
\begin{align}
S = -\frac{1}{4}\int d^5x \sqrt{-G}F_{MN}F^{MN}.
\end{align}

To proceed further, we choose an ansatz of the form\footnote{Like in the case for scalars, one might choose to work with a general ansatz $ A_t (z, \eta, \chi, \xi) $ for the gauge field component. However, one could see that this will not affect the boundary behaviour \eqref{e4.4}-\eqref{e4.5} and hence the qualitative nature of the boundary correlators as the angular dependencies are essentially decoupled from the rest of the coordinates at the level of the equations of motion which thereby allow us to fix them uniquely (like in the case for scalars) such that the equations of motion corresponding to the angular directions vanishes identically and thereby the asymptotic modes remain unaltered. For example, for our present case, the corresponding solution may be expressed as, $A_t (z, \eta, \chi, \xi)=Z(z)R(\eta)Y(\chi)e^{ik \xi}  $ where $ Y(\chi)=\sqrt[4]{|\sin ^2\chi |} \left(c_1 P_{-\frac{1}{2}}^{\frac{1}{2} \sqrt{4 k^2+1}}(\cos (\chi ))+c_2 Q_{-\frac{1}{2}}^{\frac{1}{2} \sqrt{4 k^2+1}}(\cos (\chi ))\right) $, which preserves the structure \eqref{e4.4}-\eqref{e4.5}.}
\begin{align}
A_M = (A_t (z, \eta),0,\cdots ,0)
\end{align}
which stems from the fact that we do not have any spatial currents ($ J_i $) for the boundary theory since this is a ($0+1$)d QFT. The temporal gauge field ($ A_t $) turns on a chemical potential ($ \mu $) and thereby a charge density ($ \varrho \sim J_t $) for the boundary theory. This is the case with zero frequency, $ \omega =0 $ in which case the modes $ A_\eta $ are completely decoupled from $ A_t $ near the boundary and hence do not affect the boundary charge density ($ \varrho $) that is sourced due to $ A_t $ only. Therefore, to start with, we set $ A_\eta =0 $ and these modes are important once we turn on $ \omega $.

To start with, we recast the five dimensional background \eqref{e3.12} as
\begin{align}
\label{e4.3}
ds_{2+d}^2 = ds^2_2+ z^4 f_2(z)d\eta^2+f_3 (z , \eta) d\Omega_2 (\chi , \xi)
\end{align}
where $ds^2_2$ asymptotes to an $AdS_2$ in the UV, $z\sim 0$. On the other hand, $d$ collectively denotes the rest of the spatial directions ($\eta, \chi , \xi$). Clearly, the holographic $\eta$- direction in the bulk acts like a defect to this $AdS_2$ asymptotic which in turn introduces a defect to the boundary matrix model. Like scalars, with this simplified example, our goal would be to classify vector modes that are coupled as well as decoupled from this defect, namely the $\eta$- direction.

The resulting procedure yields a set of decoupled modes in the boundary limit
\begin{align}
\label{e4.4}
&Z''+\frac{(d+2)}{z}Z'+\frac{\nu^2}{z^4}Z=0\\
&R'' +\frac{D}{\eta}R' - \nu^2 R=0
\label{e4.5}
\end{align}
where $D$ denotes the number of spatial directions of $S^2$ and we use the separation of variable, $ A_t=Z(z)R(\eta) $. Here, $\nu $ is the separation constant which carries the information of the holographic direction and thereby the defect in the ($0+1$)d QFT. This can be seen explicitly by switching-off the spatial directions, $D=0$ in which case $\nu \sim \sqrt{R''/R}$ is fixed by knowing the function along $\eta$- direction which in turn enters into \eqref{e4.4} and affects the radial profile.
\subsection{Asymptotic solutions and holographic dictionary}
Like before, we first consider the case with zero modes where we set, $\nu =0$. As explained above, these modes are ignorant about the holographic $\eta$- direction. The corresponding radial solution turns out to be
\begin{align}
\label{e4.6}
Z^{(0)}_0(z \sim 0)=\mu -\frac{c_1 z^{-d-1}}{d+1}
\end{align}
where the $ \mathcal{O}(z^0) $ term in the expansion is identified as the source or the chemical potential ($ \mu $) for the boundary theory. On the other hand, the coefficient ($c_1$) of the divergent term should be identified (modulo some proportinality constant) with the charge density ($\varrho$) in the matrix model. Clearly, this is different from what one expects for the usual $ AdS_5\times S^5 $ example \cite{Kovtun:2008kx}, which shows that we look for an alternate quantization condition for the boundary theory. One could see by setting $d=0$, that the divergence has its sole origin in the $AdS_2$ factor in \eqref{e4.3}.

The $\eta$- equation \eqref{e4.5} has a solution of the form
\begin{align}
\label{e4.7}
R_0(\eta)=r_2-\frac{r_1 \eta ^{1-D}}{D-1}.
\end{align}

Using \eqref{e4.6} and \eqref{e4.7}, the boundary charge density \cite{Kovtun:2008kx} could be formally expressed as
\begin{align}
&\varrho = \frac{\delta \mathcal{L}}{\delta(\partial_z A_t)}\Big|_{z \sim 0}\simeq 4 \pi c_1 \Big( 1-\frac{9}{8L^2}\Big)\int_0^L \eta^2 R_0 (\eta)d\eta \nonumber\\
&=4 \pi c_1 L^3 \Big(\frac{r_2}{3}-\frac{r_1}{2L}\Big).
\end{align}

Clearly, in the strict $ L \rightarrow \infty $ limit, the charge density ($ \varrho $) for the boundary theory diverges. This makes the boundary interpretation for the NATD background quite difficult and is reminiscent of what has been observed previously in the context of $ \mathcal{N}=2 $ superconformal quivers \cite{Lozano:2016kum}. Therefore, one needs to bound the holographic $ \eta $- direction in order to have a meaningful boundary interpretation. This is quite similar to the case of type IIA Gaiotto-Maldacena background, where the space is closed by putting flavour D6 branes along the holographic $ \eta $- direction \cite{Lozano:2016kum}. 

Finally, for $ \nu \neq 0 $ modes, the radial profile turns out to be
\begin{align}
\label{e4.9}
Z_\nu (z \sim 0)=-\frac{1}{4 \nu ^2 z^2}\Big( \frac{c_1 K_2\left(\frac{i\nu}{z}\right)}{\sqrt{\pi }}-32 c_2 J_2\left(\frac{\nu }{z}\right)  \Big)
\end{align}
where we set, $ d=3 $. Clearly, like in the case of scalars, the radial mode \eqref{e4.9} leads to a rapidly oscillating function and a non-renormalisable boundary action. To summarise, $ \nu \neq 0 $ modes do not have a well defined boundary interpretation.
\subsection{Boundary action and correlation functions}
Our final goal is to construct the renormalised boundary action corresponding to zero ($ \nu =0 $) modes and estimate the fields in the deep IR. We use these IR solutions to comment about the charge-charge correlation $ \mathcal{G}_{tt}(\omega) $ at finite chemical potential ($ \mu $). 

At the first place, we note down the on-shell boundary action\footnote{One might as well work out the surface term due to $ \eta $- direction which when explored close to the boundary, $ z \sim \epsilon \sim 0 $ yields a divergent contribution whose pole structure $ \sim \frac{1}{\epsilon^4} $ matches precisely to that with \eqref{e4.27} which therefore could be easily tamed by introducing an appropriate counter term.}
\begin{align}
S_B =\frac{1}{2}\int d\eta dt d\chi d\xi \sqrt{-G}G^{zz}G^{tt}F_{zt}A_t \Big |_{z\sim \epsilon}.
\end{align}

A further simplification yields
\begin{align}
\label{e4.11}
S_B \simeq- 8 \pi \Big( 1- \frac{9}{8L^2}\Big) \int d\eta dt \eta^2 z^5 \partial_z A_t A_t \Big|_{z\sim \epsilon}.
\end{align}

We propose the following Fourier decomposition for the $ U(1) $ field\footnote{We choose to work with the choice of the radial gauge, $ A_z=0 $. On the other hand, $ A_i=0 $ corresponds to the fact that there are no spatial directions to excite for the boundary theory since it lives in ($ 0+1 $)d.}
\begin{align}
\label{e4.12}
A_M = \int \frac{d\omega}{2\pi} e^{-i \omega t}A^{(\omega)}_M(z, \eta)~;~M=t, \eta
\end{align}
which yields the boundary action of the following form
\begin{align}
S_B \simeq- 8 \pi  \epsilon^5\Big( 1- \frac{9}{8L^2}\Big) \int_0^L \eta^2 d\eta \int \frac{d\omega}{2 \pi}\partial_z A_t^{(\omega)}(\epsilon) A_t^{(-\omega)}(\epsilon).
\end{align}

Using \eqref{e4.12}, one finds the following set of equations of motion\footnote{We ignore the superscript $ \omega $ for simplicity.}
\begin{align}
\label{e4.13}
&\partial_z (z^3 f_3 \partial_z A_t)+\frac{f_3}{z}(\partial^2_\eta A_t + i \omega \partial_\eta A_\eta)+\partial_\eta \Big(\frac{f_3}{z}\Big)(\partial_\eta A_t +i \omega A_\eta)=0\\
&\partial_z \Big( \frac{f_3}{z^3 f_2}\partial_z A_\eta  \Big)-\frac{f_3}{a z}(-\omega^2 A_\eta +i \omega \partial_\eta A_t)=0.
\label{e4.14}
\end{align}

The above set of equations \eqref{e4.13}-\eqref{e4.14} are difficult to solve exactly, however, they are simplified in the asymptotic $ z\sim 0 $ limit
\begin{align}
\label{e4.15}
&\partial^2_z A_t + \frac{5}{z} \partial_z A_t+\frac{1}{z^4}(\partial^2_\eta A_t + i \omega \partial_\eta A_\eta)+\frac{2}{\eta z^4}(\partial_\eta A_t +i \omega A_\eta)=0\\
&\partial^2_z A_\eta + \frac{1}{z} \partial_z A_\eta -\Big(1-\frac{3}{2L^2} \Big)(-\omega^2 A_\eta +i \omega \partial_\eta A_t)=0.
\label{e4.16}
\end{align}

In what follows, we solve the above set of equations \eqref{e4.15}-\eqref{e4.16} perturbatively in the small frequency ($ \omega \ll 1$) domain, where we propose expansions of the following form
\begin{align}
\label{e4.17}
&A_t = A_t^{(0)}+\omega A_t^{(1)}+\mathcal{O}(\omega^2)\\
&A_\eta = A_\eta^{(0)}+\omega A_\eta^{(1)}+\mathcal{O}(\omega^2).
\label{e4.18}
\end{align}

Substituting \eqref{e4.17}-\eqref{e4.18} into \eqref{e4.15}-\eqref{e4.16}, we find at leading order
\begin{align}
\label{e4.19}
&\partial^2_z A^{(0)}_t + \frac{5}{z} \partial_z A^{(0)}_t+\frac{1}{z^4}(\partial^2_\eta A^{(0)}_t +\frac{2}{\eta}\partial_\eta A^{(0)}_t)=0\\
&\partial^2_z A^{(0)}_\eta + \frac{1}{z} \partial_z A^{(0)}_\eta =0.
\label{e4.20}
\end{align}

The solution corresponding to \eqref{e4.19} could be expressed as, $ A^{(0)}_t = Z^{(0)}_0(z)R_0 (\eta) $ where the zeroth order solutions are given in \eqref{e4.6} and \eqref{e4.7}. On the other hand, \eqref{e4.20} yields a solution of the form, $ A^{(0)}_\eta= F(\eta) \log z $. Clearly, the function $ F(\eta) $ remains undetermined at $ \mathcal{O}(\omega^0) $.

We use these zeroth order solutions in \eqref{e4.15}, which yields
\begin{align}
\label{e4.21}
\partial^2_z A^{(1)}_t + \frac{5}{z} \partial_z A^{(1)}_t+\frac{1}{z^4}\partial^2_\eta A^{(1)}_t +\frac{2}{\eta z^4}(\partial_\eta A^{(1)}_t +i  \log z)=0
\end{align}
where we work with a particular choice, $ F(\eta)=1 $.

We propose a solution of the form
\begin{align}
\label{e4.23}
A^{(1)}_t (z, \eta)= -\frac{1}{\eta}Z^{(1)}_1(z)
\end{align}
which upon substitution into \eqref{e4.21}, yields the radial equation of the following form
\begin{align}
Z''^{(1)}_1 +\frac{5}{z}Z'^{(1)}_1-\frac{2i}{z^4}\log z =0.
\end{align}

The corresponding solution turns out to be
\begin{align}
\label{e4.25}
Z^{(1)}_1(z\sim 0)=Z^{(0)}_0 -\frac{i \log z}{2 z^2}.
\end{align}

We expand \eqref{e4.11} upto $ \mathcal{O}(\omega) $, which yields the renormalised boundary action
\begin{align}
S^{(ren)}_B  =S_{f}+S_{ct}
\end{align}
where the counter term could be schematically expressed as
\begin{align}
\label{e4.27}
S_{ct}=\frac{4 \pi}{\epsilon^4}(a f^{(2)}+b \omega f^{(1)})\int d\omega \varrho (\omega)\varrho (-\omega)
\end{align}
with $ a $ and $ b $ as appropriate coefficients.

The finite part of the action can be expressed as
\begin{align}
S_f =-\frac{12 }{L^3 r_2}(f^{(2)}- 2\omega f^{(1)})\int d\omega \mu (-\omega)\varrho(\omega)\equiv \int d \omega \mu (-\omega)\langle J_t(\omega)\rangle
\end{align}
where $ \langle J_t(\omega)\rangle=\frac{\delta S_f}{\delta \mu (-\omega)} \propto \varrho(\omega)$ and the coefficients $ f^{(n)} $ represent $ \eta $- integral of the form
\begin{align}
f^{(n)}=\int_0^L \eta^n R^n_0(\eta)~;~n=1,2.
\end{align}

The coupled equations \eqref{e4.13}-\eqref{e4.14} do not allow for an analytic solution away from the boundary ($ z\sim 0 $), which therefore makes it quite tricky to solve for the full two point function. However, a closer look in the deep IR ($ z \sim 1 $) reveals
\begin{align}
&\frac{3}{2 L^2} \partial^2_z A_t +\Big(\frac{3}{L^2}-2 \Big)\partial_z A_t +\frac{3}{2 L^2}(\partial^2_\eta A_t + i \omega \partial_\eta A_\eta)\approx 0\\
&\partial_z A_\eta +\frac{1}{4}(-\omega^2 A_\eta +i \omega \partial_\eta A_t) \approx 0.
\end{align}

Like before, we aim for a solution in the small frequency ($ \omega \ll 1 $) domain. The leading order $ \mathcal{O}(\omega^0) $ solutions turn out to be
\begin{align}
A_t^{(0)}\Big|_{z \sim 1}=(b_1 \eta + b_2)\Big( \frac{3 c_1 e^{\frac{2}{3} \left(2 L^2-3\right) z}}{4 L^2-6}+c_2 \Big)~;~A_\eta^{(0)}\Big|_{z \sim 1}=G(\eta)=C
\end{align}
where $ C $ is a constant.

On the other hand, the time component at $ \mathcal{O}(\omega) $ could be expressed as
\begin{align}
A_t^{(1)}\Big|_{z \sim 1}\sim A_t^{(0)}\Big|_{z \sim 1}.
\end{align}

Clearly, we see that the time component of the $ U(1) $ gauge field ($ A_t $) grows exponentially ($\sim e^{L^2 z} $) as we approach the deep IR limit, $ z \sim 1 $. On the other hand, combining \eqref{e4.6}, \eqref{e4.23} and \eqref{e4.25} one could see that the temporal component diverges as $\sim 1/\epsilon^4 $ near the asymptotic UV ($z \sim \epsilon$), where $d(=3)$ denotes the rest of the spatial directions as mentioned below \eqref{e4.3}. The function $ A_t(z) $ must therefore possesses a minima at an intermediate value $ z=z_m $.

One could therefore write down a general profile of the form
\begin{align}
\label{e4.34}
A_t (z,\eta)\sim a f(z , \eta, \omega)+b g(z, \eta , \omega).
\end{align}

The asymptotic expansions of the functions are precisely the solutions discussed above
\begin{align}
f(z , \eta , \omega)\big|_{z \sim 0}\sim -\frac{1}{\eta}(1+\omega)~;~g(z, \eta , \omega)\big|_{z \sim 0} \sim -\frac{1}{\eta}\frac{(1+\omega)}{z^4}
\end{align}
where the coefficients $ a $ and $ b $ are respectively identified as the chemical potential ($ \mu $) and the charge density ($ \varrho $) of the boundary theory.

On the other hand, in the deep IR, these functions behave as
\begin{align}
f(z , \eta , \omega)\big|_{z \sim 1}\sim \eta(1+\omega)~;~g(z, \eta , \omega)\big|_{z \sim 1} \sim \eta (1+\omega) e^{L^2 z}.
\end{align}

The ratio $ \frac{b}{a} $ determines the two point correlation ($ \mathcal{G}_{tt}(\omega) $), which can be fixed by setting
\begin{align}
A'_t (z, \eta)\Big|_{z=z_m}=0
\end{align}
where  prime denotes derivative with respect to $z$ and $ z=z_m $ is the point of inflexion.

This fixes the ratio $  \frac{b}{a} $ and yields the charge-charge correlation at finite density
\begin{align}
\label{e4.38}
\mathcal{G}_{tt}(\omega) \sim \frac{\int_0^L d\eta f'(z , \eta, \omega)|_{z=z_m}}{\int_0^L d\eta g'(z, \eta, \omega)|_{z=z_m}}
\end{align}
which is integrated over the $ \eta $- direction so that the result does not depend on the location along the holographic axis.

Before we conclude, it is worthwhile to mention that to obtain an exact ``analytic'' expression for the charge-charge correlator \eqref{e4.38} is a nontrivial task. This stems from the fact that the set of equations \eqref{e4.13}-\eqref{e4.14} are difficult to solve analytically for an intermediate range of the radial parameter $0<z<1$. For example, to our simplest imagination, one might consider the case with NATD where $L$ is set to be infinity and to begin with we set, $\omega =0$ as well. This results into the following differential equation for the temporal component of the gauge field
\begin{align}
\label{e4.39}
&\partial^2_z A_t + \Big( \frac{\left(z^2-1\right)^2 \left(3 z^2-5\right)+4 \eta ^2 \left(3 z^2-1\right) z^4}{\left(z^2-1\right)^3 z+4 \eta ^2 \left(z^2-1\right) z^5}\Big)\partial_z A_t \nonumber\\
&+\frac{1}{z^4}\partial^2_\eta A_t +\Big(\frac{2 \left(z^2-1\right)^2}{4 \eta ^3 z^8+\eta  \left(z^2-1\right)^2 z^4}\Big)\partial_\eta A_t=0
\end{align}
which does not allow an analytic solution for arbitrary $z$. To identify the form of the functions as in \eqref{e4.34} and thereby the point of inflexion ($z_m$), one might therefore look for a rigorous numerical solution of \eqref{e4.39}. This will produce a generic profile for $A_t(z,\eta)$ and would determine the correlator \eqref{e4.38} for the simplest NATD background. This is clearly beyond the scope of the present analysis and therefore we postpose this project for future investigation.
\section{Conclusions and Outlook}
Let us summarise the key outcome of the paper. The present paper deals with the holographic aspects of a class of non-Abelian T-dual (NATD) geometries \cite{Lozano:2017ole} that are obtained by performing a duality transformation along $ SU(2) $ that lives inside $ AdS_5 $. These geometries are conjectured to be dual to some ``defect'' QFT$_1 $ which can be thought of as an irrelevant deformation of the ($0+1$) dimensional matrix model preserving $\mathcal{N}=2$ SUSY. 

The purpose of the paper is to understand and build up the associated holographic dictionary for this particular class of dualities. The observable that we focus, are the correlation functions in the dual gauge theory at strong coupling. The paper presents a bulk calculation which estimates these correlation functions. In particular, we show that it is indeed possible to calculate scalar correlators and give an argument for the vector correlators of the dual QFT$_1 $.

These correlators correspond to modes in the bulk that are eventually decoupled from the holographic direction and source irrelevant operators in the dual matrix model. These operators eventually drive the theory from ($ 0+1 $) dimension to its higher dimensional parent QFT. It would be really nice to understand all these aspects from an exoplicit matrix model computation, where one starts with a particular irrelevant deformation which is dual to NATD of $ AdS_5 $.

On the dual matrix model counterpart, $\nu=0$ scalar perturbations could be thought of as (vacuum) fluctuations around fuzzy spheres which represent a particular vacuum of the PWMM. These fluctuations are sourced due to scalar operators which are typically of the form \cite{Asano:2014vba}-\cite{Asano:2014eca}
\begin{align}
\mathcal{O}(t)\sim a(t) X_i (t)+b(t)X_a (t)
\end{align}
where $N \times N$ matrices $X_i$ and $X_a$ are respectively $SO(3)$ and $SO(6)$ scalars of matrix model. 

One could therefore propose a 1- loop partition function for the dual matrix model
\begin{align}
Z_{1-loop}= Z_{PWMM}e^{\lambda\int dt \phi_0 (t)\mathcal{O}(t)}
\end{align}
where $Z_{PWMM}$ is the 1-loop partition function of the PWMM (that can be fixed using the supersymmetric localisation \cite{Asano:2012zt}) and $\lambda$ is the ``irrelevant'' coupling with mass dimension, $-n$. 

The key input of the problem comes from a proper identification of the vacuum in the dual matrix model around which one expands. One possible way to identify this vacuum is to look at the membrane solutions in the Penrose limit of $AdS_4 \times S^7$ superstar background \cite{Leblond:2001gn}, that is reduced down to type IIA NATD geometry following a suitable dimensional reduction \cite{Lozano:2017ole}. Once this vacuum is properly identified, one could think of an expansion as mass deformation of the PWMM that is controlled by the coupling constant $\lambda$ carrying negative mass dimension.

\subsection*{Acknowledgments}
It's a pleasure to thank Carlos Nunez for his careful reading of the draft and making several useful comments that has improved the presentation. The author is indebted to the authorities of IIT Roorkee for their unconditional support towards researches in basic sciences. The author would like to acknowledge the Mathematical Research Impact Centric Support (MATRICS) grant (MTR/2023/000005) received from SERB, India. The author also acknowledges The Royal Society, UK for financial assistance.



\begin{thebibliography}{99}
\bibitem{Maldacena:1997re}
J.~M.~Maldacena,
``The Large N limit of superconformal field theories and supergravity,''
Adv. Theor. Math. Phys. \textbf{2}, 231-252 (1998)
doi:10.4310/ATMP.1998.v2.n2.a1
[arXiv:hep-th/9711200 [hep-th]].

\bibitem{Aharony:1999ti}
O.~Aharony, S.~S.~Gubser, J.~M.~Maldacena, H.~Ooguri and Y.~Oz,
``Large N field theories, string theory and gravity,''
Phys. Rept. \textbf{323}, 183-386 (2000)
doi:10.1016/S0370-1573(99)00083-6
[arXiv:hep-th/9905111 [hep-th]].

\bibitem{Ramallo:2013bua}
A.~V.~Ramallo,
``Introduction to the AdS/CFT correspondence,''
Springer Proc. Phys. \textbf{161}, 411-474 (2015)
doi:10.1007/978-3-319-12238-0\_10
[arXiv:1310.4319 [hep-th]].

\bibitem{Natsuume:2014sfa}
M.~Natsuume,
``AdS/CFT Duality User Guide,''
Lect. Notes Phys. \textbf{903}, pp.1-294 (2015)
doi:10.1007/978-4-431-55441-7
[arXiv:1409.3575 [hep-th]].

\bibitem{Gubser:1998bc}
S.~S.~Gubser, I.~R.~Klebanov and A.~M.~Polyakov,
``Gauge theory correlators from noncritical string theory,''
Phys. Lett. B \textbf{428}, 105-114 (1998)
doi:10.1016/S0370-2693(98)00377-3
[arXiv:hep-th/9802109 [hep-th]].

\bibitem{Witten:1998qj}
E.~Witten,
``Anti-de Sitter space and holography,''
Adv. Theor. Math. Phys. \textbf{2}, 253-291 (1998)
doi:10.4310/ATMP.1998.v2.n2.a2
[arXiv:hep-th/9802150 [hep-th]].

\bibitem{Son:2002sd}
D.~T.~Son and A.~O.~Starinets,
``Minkowski space correlators in AdS / CFT correspondence: Recipe and applications,''
JHEP \textbf{09}, 042 (2002)
doi:10.1088/1126-6708/2002/09/042
[arXiv:hep-th/0205051 [hep-th]].

\bibitem{Skenderis:2002wp}
K.~Skenderis,
``Lecture notes on holographic renormalization,''
Class. Quant. Grav. \textbf{19}, 5849-5876 (2002)
doi:10.1088/0264-9381/19/22/306
[arXiv:hep-th/0209067 [hep-th]].

\bibitem{deHaro:2000vlm}
S.~de Haro, S.~N.~Solodukhin and K.~Skenderis,
``Holographic reconstruction of space-time and renormalization in the AdS / CFT correspondence,''
Commun. Math. Phys. \textbf{217}, 595-622 (2001)
doi:10.1007/s002200100381
[arXiv:hep-th/0002230 [hep-th]].

\bibitem{Kovtun:2008kx}
P.~Kovtun and A.~Ritz,
``Universal conductivity and central charges,''
Phys. Rev. D \textbf{78}, 066009 (2008)
doi:10.1103/PhysRevD.78.066009
[arXiv:0806.0110 [hep-th]].

\bibitem{Lozano:2017ole}
Y.~Lozano, C.~Nunez and S.~Zacarias,
``BMN Vacua, Superstars and Non-Abelian T-duality,''
JHEP \textbf{09}, 008 (2017)
doi:10.1007/JHEP09(2017)008
[arXiv:1703.00417 [hep-th]].

\bibitem{delaOssa:1992vci}
X.~C.~de la Ossa and F.~Quevedo,
``Duality symmetries from nonAbelian isometries in string theory,''
Nucl. Phys. B \textbf{403}, 377-394 (1993)
doi:10.1016/0550-3213(93)90041-M
[arXiv:hep-th/9210021 [hep-th]].

\bibitem{Alvarez:1993qi}
E.~Alvarez, L.~Alvarez-Gaume, J.~L.~F.~Barbon and Y.~Lozano,
``Some global aspects of duality in string theory,''
Nucl. Phys. B \textbf{415}, 71-100 (1994)
doi:10.1016/0550-3213(94)90067-1
[arXiv:hep-th/9309039 [hep-th]].

\bibitem{Alvarez:1994np}
E.~Alvarez, L.~Alvarez-Gaume and Y.~Lozano,
``On nonAbelian duality,''
Nucl. Phys. B \textbf{424}, 155-183 (1994)
doi:10.1016/0550-3213(94)90093-0
[arXiv:hep-th/9403155 [hep-th]].

\bibitem{Sfetsos:2010uq}
K.~Sfetsos and D.~C.~Thompson,
``On non-abelian T-dual geometries with Ramond fluxes,''
Nucl. Phys. B \textbf{846}, 21-42 (2011)
doi:10.1016/j.nuclphysb.2010.12.013
[arXiv:1012.1320 [hep-th]].

\bibitem{Lozano:2011kb}
Y.~Lozano, E.~O Colgain, K.~Sfetsos and D.~C.~Thompson,
``Non-abelian T-duality, Ramond Fields and Coset Geometries,''
JHEP \textbf{06}, 106 (2011)
doi:10.1007/JHEP06(2011)106
[arXiv:1104.5196 [hep-th]].

\bibitem{Itsios:2012dc}
G.~Itsios, Y.~Lozano, E.~O Colgain and K.~Sfetsos,
``Non-Abelian T-duality and consistent truncations in type-II supergravity,''
JHEP \textbf{08}, 132 (2012)
doi:10.1007/JHEP08(2012)132
[arXiv:1205.2274 [hep-th]].

\bibitem{Itsios:2012zv}
G.~Itsios, C.~Nunez, K.~Sfetsos and D.~C.~Thompson,
``On Non-Abelian T-Duality and new N=1 backgrounds,''
Phys. Lett. B \textbf{721}, 342-346 (2013)
doi:10.1016/j.physletb.2013.03.033
[arXiv:1212.4840 [hep-th]].

\bibitem{Itsios:2013wd}
G.~Itsios, C.~Nunez, K.~Sfetsos and D.~C.~Thompson,
``Non-Abelian T-duality and the AdS/CFT correspondence:new N=1 backgrounds,''
Nucl. Phys. B \textbf{873}, 1-64 (2013)
doi:10.1016/j.nuclphysb.2013.04.004
[arXiv:1301.6755 [hep-th]].

\bibitem{ReidEdwards:2010qs}
R.~A.~Reid-Edwards and B.~Stefanski, jr.,
``On Type IIA geometries dual to N = 2 SCFTs,''
Nucl. Phys. B \textbf{849}, 549-572 (2011)
doi:10.1016/j.nuclphysb.2011.04.002
[arXiv:1011.0216 [hep-th]].

\bibitem{Aharony:2012tz}
O.~Aharony, L.~Berdichevsky and M.~Berkooz,
``4d N=2 superconformal linear quivers with type IIA duals,''
JHEP \textbf{08}, 131 (2012)
doi:10.1007/JHEP08(2012)131
[arXiv:1206.5916 [hep-th]].

\bibitem{Barranco:2013fza}
A.~Barranco, J.~Gaillard, N.~T.~Macpherson, C.~N\'u\~nez and D.~C.~Thompson,
``G-structures and Flavouring non-Abelian T-duality,''
JHEP \textbf{08}, 018 (2013)
doi:10.1007/JHEP08(2013)018
[arXiv:1305.7229 [hep-th]].

\bibitem{Macpherson:2014eza}
N.~T.~Macpherson, C.~N\'u\~nez, L.~A.~Pando Zayas, V.~G.~J.~Rodgers and C.~A.~Whiting,
``Type IIB supergravity solutions with AdS$_{5}$ from Abelian and non-Abelian T dualities,''
JHEP \textbf{02}, 040 (2015)
doi:10.1007/JHEP02(2015)040
[arXiv:1410.2650 [hep-th]].

\bibitem{Lozano:2016kum}
Y.~Lozano and C.~N\'u\~nez,
``Field theory aspects of non-Abelian T-duality and $ \mathcal{N}  =$ 2 linear quivers,''
JHEP \textbf{05}, 107 (2016)
doi:10.1007/JHEP05(2016)107
[arXiv:1603.04440 [hep-th]].

\bibitem{Roychowdhury:2021eas}
D.~Roychowdhury,
``Fragmentation and defragmentation of strings in type IIA and their holographic duals,''
JHEP \textbf{08}, 030 (2021)
doi:10.1007/JHEP08(2021)030
[arXiv:2104.11953 [hep-th]].

\bibitem{Nunez:2018qcj}
C.~Nunez, D.~Roychowdhury and D.~C.~Thompson,
``Integrability and non-integrability in $ \mathcal{N}=2 $ SCFTs and their holographic backgrounds,''
JHEP \textbf{07}, 044 (2018)
doi:10.1007/JHEP07(2018)044
[arXiv:1804.08621 [hep-th]].

\bibitem{Caceres:2014uoa}
E.~Caceres, N.~T.~Macpherson and C.~N\'u\~nez,
``New Type IIB Backgrounds and Aspects of Their Field Theory Duals,''
JHEP \textbf{08}, 107 (2014)
doi:10.1007/JHEP08(2014)107
[arXiv:1402.3294 [hep-th]].

\bibitem{Araujo:2015dba}
T.~R.~Araujo and H.~Nastase,
``Non-Abelian T-duality for nonrelativistic holographic duals,''
JHEP \textbf{11}, 203 (2015)
doi:10.1007/JHEP11(2015)203
[arXiv:1508.06568 [hep-th]].

\bibitem{Itsios:2019yzp}
G.~Itsios, J.~M.~Pen\'\i{}n and S.~Zacar\'\i{}as,
``Spin-2 excitations in Gaiotto-Maldacena solutions,''
JHEP \textbf{10}, 231 (2019)
doi:10.1007/JHEP10(2019)231
[arXiv:1903.11613 [hep-th]].

\bibitem{Nunez:2019gbg}
C.~N\'u\~nez, D.~Roychowdhury, S.~Speziali and S.~Zacar\'\i{}as,
``Holographic aspects of four dimensional ${\cal N }=2$ SCFTs and their marginal deformations,''
Nucl. Phys. B \textbf{943}, 114617 (2019)
doi:10.1016/j.nuclphysb.2019.114617
[arXiv:1901.02888 [hep-th]].

\bibitem{vanGorsel:2017goj}
J.~van Gorsel and S.~Zacar\'\i{}as,
``A Type IIB Matrix Model via non-Abelian T-dualities,''
JHEP \textbf{12}, 101 (2017)
doi:10.1007/JHEP12(2017)101
[arXiv:1711.03419 [hep-th]].

\bibitem{Roychowdhury:2021unp}
D.~Roychowdhury,
``Matrix models and non-Abelian T dual of $AdS_5 \times S^5 $,''
doi:10.1002/prop.202300146
[arXiv:2110.05395 [hep-th]].

\bibitem{Roychowdhury:2023lxk}
S.~Roychowdhury and D.~Roychowdhury,
``Spin 2 spectrum for marginal deformations of 4d $ \mathcal{N} $ = 2 SCFTs,''
JHEP \textbf{03}, 083 (2023)
doi:10.1007/JHEP03(2023)083
[arXiv:2301.12757 [hep-th]].

\bibitem{Gaiotto:2009gz}
D.~Gaiotto and J.~Maldacena,
``The Gravity duals of N=2 superconformal field theories,''
JHEP \textbf{10}, 189 (2012)
doi:10.1007/JHEP10(2012)189
[arXiv:0904.4466 [hep-th]].

\bibitem{Gaiotto:2009we}
D.~Gaiotto,
``N=2 dualities,''
JHEP \textbf{08}, 034 (2012)
doi:10.1007/JHEP08(2012)034
[arXiv:0904.2715 [hep-th]].

\bibitem{Berenstein:2002jq}
D.~E.~Berenstein, J.~M.~Maldacena and H.~S.~Nastase,
``Strings in flat space and pp waves from N=4 superYang-Mills,''
JHEP \textbf{04}, 013 (2002)
doi:10.1088/1126-6708/2002/04/013
[arXiv:hep-th/0202021 [hep-th]].

\bibitem{Maldacena:2002rb}
J.~M.~Maldacena, M.~M.~Sheikh-Jabbari and M.~Van Raamsdonk,
``Transverse five-branes in matrix theory,''
JHEP \textbf{01}, 038 (2003)
doi:10.1088/1126-6708/2003/01/038
[arXiv:hep-th/0211139 [hep-th]].

\bibitem{Banks:1996vh}
T.~Banks, W.~Fischler, S.~H.~Shenker and L.~Susskind,
``M theory as a matrix model: A Conjecture,''
Phys. Rev. D \textbf{55}, 5112-5128 (1997)
doi:10.1103/PhysRevD.55.5112
[arXiv:hep-th/9610043 [hep-th]].

\bibitem{Lin:2004nb}
H.~Lin, O.~Lunin and J.~M.~Maldacena,
``Bubbling AdS space and 1/2 BPS geometries,''
JHEP \textbf{10}, 025 (2004)
doi:10.1088/1126-6708/2004/10/025
[arXiv:hep-th/0409174 [hep-th]].

\bibitem{Lin:2004kw}
H.~Lin,
``The Supergravity dual of the BMN matrix model,''
JHEP \textbf{12}, 001 (2004)
doi:10.1088/1126-6708/2004/12/001
[arXiv:hep-th/0407250 [hep-th]].

\bibitem{Lin:2005nh}
H.~Lin and J.~M.~Maldacena,
``Fivebranes from gauge theory,''
Phys. Rev. D \textbf{74}, 084014 (2006)
doi:10.1103/PhysRevD.74.084014
[arXiv:hep-th/0509235 [hep-th]].

\bibitem{book}Abramowitz, Milton; Stegun, Irene A. (1964). "Section 6.4". Handbook of Mathematical Functions. New York: Dover Publications. ISBN 978-0-486-61272-0.

\bibitem{Asano:2014vba}
Y.~Asano, G.~Ishiki, T.~Okada and S.~Shimasaki,
``Emergent bubbling geometries in the plane wave matrix model,''
JHEP \textbf{05}, 075 (2014)
doi:10.1007/JHEP05(2014)075
[arXiv:1401.5079 [hep-th]].

\bibitem{Asano:2014eca}
Y.~Asano, G.~Ishiki and S.~Shimasaki,
``Emergent bubbling geometries in gauge theories with SU(2|4) symmetry,''
JHEP \textbf{09}, 137 (2014)
doi:10.1007/JHEP09(2014)137
[arXiv:1406.1337 [hep-th]].

\bibitem{Asano:2012zt}
Y.~Asano, G.~Ishiki, T.~Okada and S.~Shimasaki,
``Exact results for perturbative partition functions of theories with SU(2|4) symmetry,''
JHEP \textbf{02}, 148 (2013)
doi:10.1007/JHEP02(2013)148
[arXiv:1211.0364 [hep-th]].

\bibitem{Leblond:2001gn}
F.~Leblond, R.~C.~Myers and D.~C.~Page,
``Superstars and giant gravitons in M theory,''
JHEP \textbf{01}, 026 (2002)
doi:10.1088/1126-6708/2002/01/026
[arXiv:hep-th/0111178 [hep-th]].


\end{thebibliography}
\end{document}